\documentclass[lettersize,journal]{IEEEtran}
\usepackage{amsmath,amsfonts}
\usepackage{algorithmic}
\usepackage{array}
\usepackage{textcomp}
\usepackage{stfloats}
\usepackage{url}
\usepackage{verbatim}
\usepackage{graphicx}
\usepackage{subfigure}
\usepackage{color}
\usepackage[linesnumbered,ruled,vlined]{algorithm2e}
\def\BibTeX{{\rm B\kern-.05em{\sc i\kern-.025em b}\kern-.08em
    T\kern-.1667em\lower.7ex\hbox{E}\kern-.125emX}}
\usepackage{balance}
\begin{document}

\title{Interpretable Nonroutine Network Traffic Prediction with a Case Study}

\author{Liangzhi Wang, Haoyuan Zhu, Jiliang Zhang, \emph{Senior Member, IEEE}, Zitian Zhang, \\and Jie Zhang, \emph{Senior Member, IEEE}

\thanks{Liangzhi Wang and Haoyuan Zhu are with the Department of Electronic and Electrical Engineering, the University of Sheffield, Sheffield, S10 2TN, UK. (e-mail: lwang85@sheffield.ac.uk, hzhu51@sheffield.ac.uk).

Jiliang Zhang is with College of Information Science and Engineering, Northeastern University, Shenyang, China, and with National Mobile Communications Research Laboratory, Southeast University, China. (e-mail: zhangjiliang1@mail.neu.edu.cn).

Zitian Zhang is with School of Information and Electronic Engineering, Zhejiang Gongshang University, Hangzhou, China. (e-mail: zitian.zhang@mail.zjgsu.edu.cn).

Jie Zhang is with the Department of Electronic and Electrical Engineering, the University of Sheffield, Sheffield, S10 2TN, UK, and also with Ranplan Wireless Network Design Ltd., Cambridge, CB23 3UY, UK. (e-mail: jie.zhang@sheffield.ac.uk).

This work is supported by the open research fund of National Mobile Communications Research Laboratory, Southeast University (2024D09).
}
}

\markboth{IEEE}%
{How to Use the IEEEtran \LaTeX \ Templates}

\maketitle

\begin{abstract}

This paper pioneers a nonroutine network traffic prediction (NNTP) method 
to prospectively provide a theoretical basis for avoiding large-scale network disruption by accurately predicting bursty traffic.
Certain events that impact user behavior subsequently trigger nonroutine traffic, which significantly constrains the performance of network traffic prediction (NTP) models.
By analyzing nonroutine traffic and the corresponding events, the NNTP method is pioneered to construct interpretable NTP model.
Based on the real-world traffic data, the network traffic generated during soccer games serves as a case study to validate the performance of the NNTP method. 
The numerical results indicate that our prediction closely fits the traffic pattern. 
In comparison to existing researches, the NNTP method is at the forefront of finding a balance among interpretability, accuracy, and computational complexity.
\end{abstract}


\section{Introduction}

\IEEEPARstart 
{A}{ccording} to the Ericsson's report, the monthly global mobile traffic has reached a staggering 143 EB by the end of third quarter in 2023, and is forecast to rise to 563 EB by the end of 2029 \cite{Ericsson}.
Such a large volume of global mobile traffic poses a severe challenge to network efficiency. As one of the most important technologies regarding network resource allocation \cite{W. Jiang}, network traffic prediction (NTP) has garnered widespread attention within the academic community. 

This paper discovers the phenomenon that certain events can bring about significant changes in cellular network traffic by impacting user behavior. This kind of traffic, referred to as nonroutine traffic, poses a great challenge to the state-of-the-art NTP models in terms of prediction accuracy, computational efficiency, and interpretability. This problem caused by nonroutine traffic, overlooked in academia, carries significant potential consequences. The degradation in NTP performance is anticipated to result in a decline in the quality of service (QoS), consequently leading to a deterioration in customer satisfaction \cite{customer satisfaction} and the reputation of the operator \cite{D. Jhamb}. Ultimately, this may culminate in severe repercussions, including compromised future profitability \cite{D. Jhamb}.

The state-of-the-art NTP models exhibit inherent limitations, and experimental results in this paper indicate that these limitations are exacerbated when it encounters nonroutine traffic data.
The contemporary state-of-the-art researches focus on historical network traffic data, while neglect the influence of user behaviour on traffic pattern. These researches can be chiefly divided into three categories namely, the statistics-based model, the shallow-learning-based model and deep-learning-based model \cite{Cellular Traffic Prediction}\cite{Y. Wang}. The statistics-based and shallow-learning-based models have limited learning capacity \cite{dmTP}. The nonroutine traffic data will further hinder their performance. Although the deep-learning-based model could understand the complex temporal-spatial correlations \cite{Fuyou}, it is still difficult to extract the traffic pattern accurately when confronted with nonroutine traffic data. It is because the nonroutine data only takes up a relatively small proportion of the overall data. Meanwhile, it requires not only a large amount of nonroutine traffic data, but also an increase in the parameters and complexity of the model. This can further increase the difficulty of hyper-parameters' selection and the risk of overfitting, as well as introduce longer computation time. Moreover, Machine Learning (ML)-based models have poor interpretability for nonroutine traffic data, because its parameters do not have practical significance \cite{Liangzhi}.
Consequently, the state-of-the-art NTP models perform poorly in the presence of nonroutine traffic data.

To achieve a leap from 0 to 1 in the context of nonroutine network traffic, this paper pioneers a novel nonroutine network traffic prediction (NNTP) method. Specifically, using the the real-world traffic data generated during soccer games as a case study, this paper initially analyzes the underlying causes of the nonroutine traffic. It subsequently reveals the correlation between user behavior and traffic pattern. 
Finally, it formulates a dedicated model, referred to as the soccer game nonroutine network traffic prediction (SG-NNTP) model, for such nonroutine events.
The model constructed based on the NNTP method is analytical and interpretable. In addition, numerical results show that the NNTP method performs excellently in prediction accuracy and computational efficiency.

The main contributions of this paper are summarized as follows:
\begin{enumerate}
\item[1)]
This paper takes the lead in systematically analyzing and researching nonroutine traffic, i.e. network traffic caused by nonroutine event which differs significantly from regular traffic patterns.
\item[2)]
Based on the analysis of nonroutine traffic, this paper pioneers the NNTP method to construct the  NTP model. The NNTP method is an important inspiration for future research on nonroutine traffic. 
\item[3)]
Following the NNTP method, this paper formulates the SG-NNTP model as a case study. Compared with benchmark models which do not take nonroutine traffic into consideration, the NNTP method improves the prediction accuracy, both in multi-step and single-step prediction mode.  What is more, the NNTP method decreases the elapsed time and improves the computational efficiency a lot. In addition, the NNTP method has outstanding interpretability and is easy to migrate to similar situations.
\end{enumerate}

The rest of the paper is structured as follows. In Section II, this paper reviews the related works on NTP field. Section III offers the definition and categorization of nonroutine traffic. It subsequently introduces the key points of the NNTP method and formulates the SG-NNTP model as a case study. 
In Section IV, the predictions of the SG-NNTP model and benchmark models are performed, and the performance of these models is evaluated in both multi-step and single-step prediction mode.
Finally, we conclude this work in Section V.

\section{Related Works}
In the past few years, academia has made great efforts in the NTP field. 
In traditional statistics-based models, historical network traffic data is fitted into some statistics or probability distributions to extract traffic patterns and attributes, such as the $\alpha$-stable model, the Autoregressive Moving Average (ARMA) model, the Autoregressive Integrated Moving Average (ARIMA) model, the seasonal ARIMA (SARIMA) model, etc. 
The authors in \cite{X. Ge} proposed the $\alpha$-stable model to predict traffic fluctuations by using the traffic loads' significant self-similarity.
As the representation for traditional statistics-based models, the ARMA and ARIMA model are often used to extract linear features from historical traffic data \cite{I. Loh}. 
The ARMA model is suitable for stationary sequence. 
L. Tang \textit {et al.} employed the ARMA model to predict the future load state of virtual networks, and then based on the prediction results, proposed a dynamic resource allocation scheme for virtual networks \cite{yuan5}.
While the ARIMA model fits non-stationary sequence well by adding a difference-stationary process \cite{yuan6}. 
In \cite{yuan7}, the ARIMA model was used to predict the normal traffic in the next minute to identify DoS and DDoS attacks. 
As an extension of the ARIMA model, the SARIMA model is more concerned with the seasonal variations of time series \cite{Y. Shu}. Besides, some nonlinear models were proposed for nonlinear features, such as the generalized auto-regressive conditional heteroskedasticity (GARCH) model \cite{yuan8}. 
In addition to the models mentioned above, the covariance function \cite{covariance function}, the ON-OFF model \cite{ON-OFF model}, and the Holt-Winter's exponential smoothing model \cite{Holt-Winter} were proposed to suit the temporal and/or spatial features of network traffic data.
Although traditional statistics-based models have the advantage of low computational complexity, in general, they do not perform very well in prediction accuracy in comparison to ML-based models \cite{Amin Azari}.
 
Given the exponential growth of traffic data and the rapid development of ML, data driven ML-based NTP models are increasingly emerging and have gained significant attention among researchers. 
Multiple activation functions enable the ML-based models to comprehend complicated nonlinear features. ML-based models can be further divided into shallow-learning-based and deep-learning-based models [6-9]. 
The shallow-learning algorithms, like support vector regression (SVR) \cite{SVR} and linear regression \cite{LR}, have been firstly introduced in NTP field and achieve some progress. 
Nevertheless, their limited learning capability do not align with complex characteristics of network traffic. They may still achieve low prediction accuracy even though there are sufficient learning samples.

In recent years, the advancement of deep learning techniques has propelled ML-based models towards deep learning, consequently boosting their learning capabilities. Deep-learning-based models represented by deep neural networks (DNNs) perform well in terms of prediction accuracy.
J. Zhou \textit {et al.} in \cite{Jian Zhou} proposed a multi-scale deep echo-state network (ESN) based NTP model to learn the characteristics of the network traffic associated with Internet of Things (IoT) application, such as multiscale, nonlinearity, and scale dependence. 
To further enhance the accuracy, Z. Tian \textit {et al.} proposed a combination NTP model based on bidirectional LSTM network to forecast the product function components and a residual decomposed from network traffic by local mean decomposition method \cite{Z. Tian}.
S. Wang \textit {et al.} in \cite{S. Wang} proposed a multi-task learning based deep-learning network, combining LSTM and CNN, to improve the performance of NTP models under the influence of large amount of irregular fluctuations in IoT network traffic.
In \cite{Cellular Traffic Prediction}, a spatio-temporal feature-learning model was proposed to capture non-local spatial correlations and learn the correlation of different time-grained features.
In order to extract the spatio-temporal dependence and characteristics of cellular traffic and thus improve the prediction accuracy, Z. Wang \textit {et al.} proposed  a time-series similarity-based graph attention network \cite{Z. Wang}.
Deep-learning-based NTP models have made great efforts in terms of prediction accuracy, but the increase in accuracy has leaded to a concomitant increase in model complexity and an elevated amount of model parameters. 
As a result, hyperparameter selection is both complex and time-consuming, which reduces computational efficiency. Moreover, black-box model significantly weakens interpretability.

\section{The Proposed NNTP Method with a Case Study}
This section proposes the NNTP method to enhance the performance of the NTP model in the presence of nonroutine traffic data. Following this method, we formulate the SG-NNTP model with the real-world traffic data gathered during soccer games.

\subsection{Analysis of Nonroutine Traffic}
In a specific geographical area, the daily activities of local users typically exhibit a high degree of cyclical and repetitive pattern. 
The network traffic that determined by user behavior demonstrates a similar variation trend.
Therefore, the user-behaviour-based network traffic prediction (UBB NTP) model can quickly and accurately capture daily traffic pattern \cite{Liangzhi}.

However, the incidence of nonroutine events in the region may influence user behavior or the quantity of user, subsequently exerting a substantial impact on the traffic pattern.
When the region hosts significant events, such as sports games and concerts, it tends to draw a considerable influx of short-term users. 
In this context, short-term users refer to individuals who arrive in this region specifically for the event and stay there during its duration.
The traffic generated by these short-term users is significantly different from the daily traffic. This type of nonroutine traffic is defined as additive nonroutine traffic. As the name suggests, in this case, the overall traffic in the area can be regarded as a superposition of the daily traffic generated by resident users and the nonroutine traffic generated by short-term users.

\begin{figure}[!t]
\centering 
\includegraphics[width=3.4in]{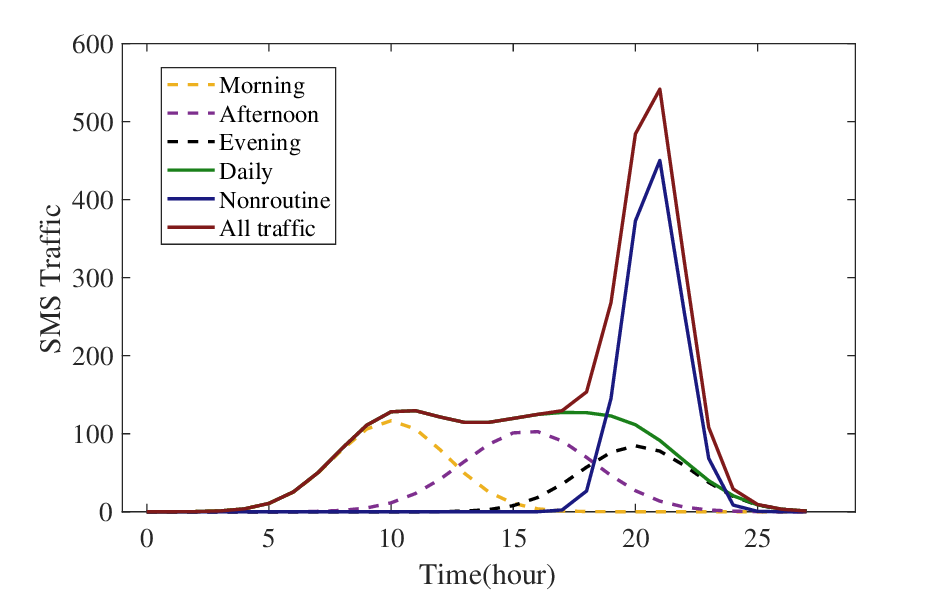}
\caption{The daily traffic and the additive nonroutine traffic.}
\label{Daily and additive traffic}
\end{figure}

\subsection{The Proposed NNTP Method}
Based on the analysis of nonroutine traffic, this paper proposes the NNTP method. In simple terms, the NNTP method involves analyzing and summarizing the traffic data corresponding to the nonroutine events, and formulating specific NTP models for similar events. 

Many events in the daily lives of users could make the daily traffic pattern change abnormally. The occurrences of these events are often scheduled rather than completely random and unexpected, such as fairs, ball games, concerts, carnivals, and so on. Some of the information related to these scheduled events is available in advance, referred to as advanced information, such as the commencement time, the event duration, the expected attendance or ticket sales, and the type of the event. 
Relying solely on historical traffic data to acquire these advanced information is difficult and demanding.

It would be a shortcut to directly utilize this easily accessible advanced information to construct NTP models corresponding to these nonroutine events. Meanwhile, with today's data explosion, effective multi-source data application is becoming an essential research. Hence, the efficient utilization of multi-source data is also one of the innovations of the NNTP method. Specifically, this paper decomposes the total traffic during the nonroutine event into a superposition of the daily traffic and the additive nonroutine traffic.
As shown in Fig. \ref{Daily and additive traffic}, the solid green line depicts the daily traffic, the solid blue one depicts the additive nonroutine traffic, and the solid red one represents the total traffic.
It is worth noting that the horizontal axis in Fig. \ref{Daily and additive traffic} represents time in hours.  The values of the horizontal axis should not exceed 23; any values greater than or equal to 24 should be attributed to the next day. However, for the sake of axis continuity, a representation beyond 23 has been adopted in this paper.

\begin{figure}[!t]
\centering 
\includegraphics[width=3.4in]{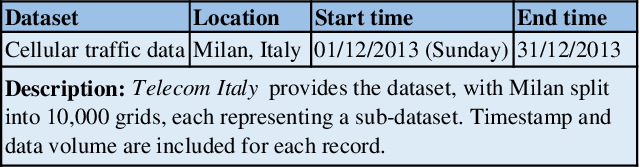}
\caption{Cellular network traffic data in Milan published by Telecom Italia \cite{G. Barlacchi}.}
\label{dataset information}
\end{figure}

\begin{figure*}[!t]
\centering 
\includegraphics[width=5.75in]{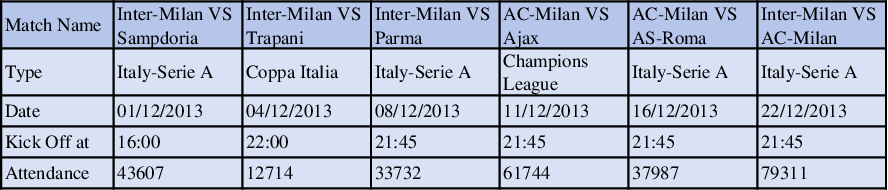} 
\caption{The information of the soccer games hosted by the San Siro Stadium in December 2013 \cite{Federico Botta}.}
\label{soccer games information}
\end{figure*}

\begin{algorithm}[t]
\label{algorithm2}
	\renewcommand{\algorithmicrequire}{\textbf{Input:}}
	\renewcommand{\algorithmicensure}{\textbf{Output:}}
	\caption{the Proposed NNTP Method}
	\begin{algorithmic}[1]
	\REQUIRE $\textbf{S},~\textbf{AI} = [\textbf{h},~\textbf{d},~\textbf{g},~\textbf{m}]$ 
        \STATE Dividing the historical traffic data $\textbf{S}$ into two parts: $\textbf{S}_\alpha$ and $\textbf{S}_\beta$ 
        \STATE Construct the Module.UBBNTP based on $\textbf{S}_\alpha$
        \STATE $\textbf{S}_{\beta,\rm daily}$ = Module.UBBNTP$\left( \textbf{h},~\textbf{d} \right)$ \ \% Predict the daily traffic component in $\textbf{S}_{\beta}$
        \STATE $\textbf{S}_{\beta,\rm nonroutine}$ = $\textbf{S}_{\beta} - \textbf{S}_{\beta,\rm daily}$
		\FOR{$i = 1: \text{lenth}\left(\textbf{AI}\right)$}
			\STATE Get $\textbf{AI}[i]$
			\STATE Infer the connection $\textbf{C}[i]$ between $\textbf{S}_{\beta,\rm nonroutine}$ and $\textbf{AI}[i]$
		\ENDFOR
		\ENSURE Construct the specific NNTP model based on $\textbf{C}$ and $\textbf{S}_{\beta,\rm nonroutine}$
	\end{algorithmic}  
\end{algorithm}

Next, the pseudo-code \textbf{Algorithm 1} is used for introducing the proposed NNTP method. In \textbf{Algorithm 1}, $\textbf{S}$ represents the historical traffic data. $\textbf{S}$ is categorized into two groups, i.e. $\textbf{S}_\alpha$ and $\textbf{S}_\beta$, depending on the occurrence of nonroutine events. $\textbf{S}_\alpha$ represents the total traffic data generated on days when the nonroutine events do not occur, while $\textbf{S}_\beta$ represents the total traffic data generated on days when the nonroutine events take place. Then, we construct the UBB NTP model that relies on $\textbf{S}_\alpha$ and forecast the the daily traffic component $\textbf{S}_{\beta,\rm daily}$ in $\textbf{S}_\beta$. Thus, the nonroutine traffic component $\textbf{S}_{\beta,\rm nonroutine}$ in $\textbf{S}_\beta$ can be obtained.
$\textbf{AI}$ represents the advanced information, including but not limited to the commencement time $\textbf{h}$, the duration $\textbf{d}$, the type $\textbf{g}$, and the attendance $\textbf{m}$ of the nonroutine events. By conducting a thorough analysis of nonroutine traffic, we can extract the relationship $\textbf{C}$ between $\textbf{S}_{\beta,\rm nonroutine}$ and $\textbf{AI}$. Finally, the specific NNTP model can be constructed. Next, this paper will detail the process through a case study.

\subsection{Dataset for Nonroutine Traffic}
This paper adopts the real-world network traffic data, as shown in Fig. \ref{dataset information}, published by Telecom Italia, a large European telecommunications service operator \cite{G. Barlacchi}. In the spatial dimension, Milan city is covered by 10,000 grids of size $235\times235$ meters. Each data point within every grid represents the cellular traffic data generated by local users during the time interval between two consecutive timestamps.

By searching the grids surrounding the G.MEAZZA SAN SIRO, we obtained traffic data in the vicinity of the San Siro stadium. The time granularity is set at one hour, which means that there are 24 data samples each day. Then, this paper researches the information of the soccer game hosted by the San Siro Stadium in December 2013, as shown in Fig. \ref{soccer games information}.

\begin{figure}[!t]
\centering 
\includegraphics[width=3.4in]{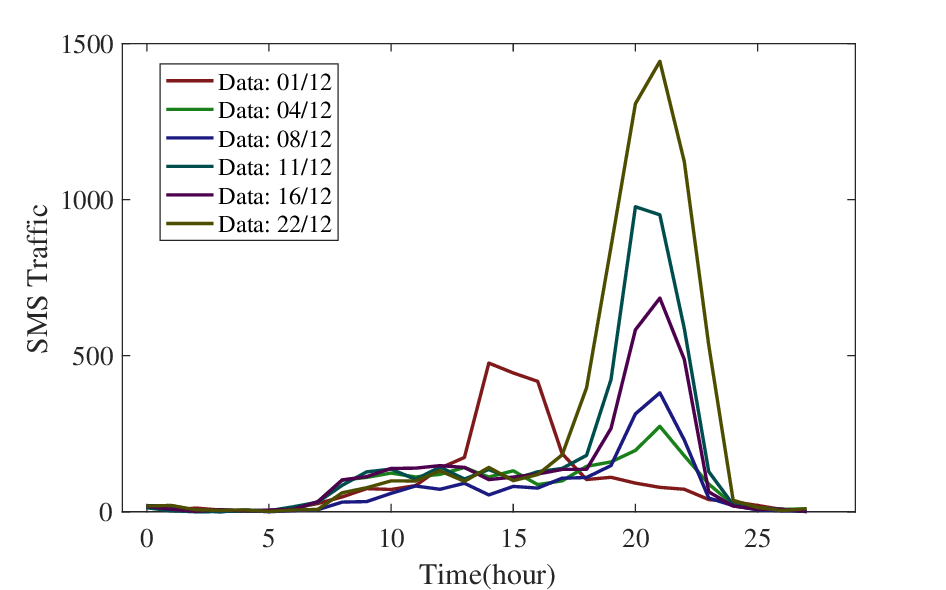}
\caption{SMS data for the soccer games held at the San Siro stadium in December.}
\label{SMS data for the soccer games}
\end{figure}

Authors in \cite{Federico Botta} demonstrates that the traffic data, including Short Message Service (SMS), calls, internet, etc., is directly related to the soccer games and contains similar nonroutine traffic pattern. Take SMS data as an example, Fig. \ref{SMS data for the soccer games} plots the curves of the traffic data corresponding to all of the soccer games in Fig. \ref{soccer games information}. 
Visual inspection reveals an alignment between the time of peaks and the period of the soccer game, which is consistent with the conclusion given by F. Botta \textit {et al.} in \cite{Federico Botta}.

\subsection{The Case Study: SG-NNTP Model}
Based on the above analysis, the traffic brought by soccer games belongs to the additive nonroutine traffic. 
To extract the nonroutine component, the initial step is to obtain the daily traffic component with the UBBNTP method. There are three dedicated daily traffic models designed for weekday, Saturday, and Sunday, respectively \cite{Liangzhi}.
As shown in Fig. \ref{Daily and additive traffic}, any of the daily traffic model represented by solid green line is a superposition of yellow, purple, and black dashed lines which represent the traffic components with the morning, afternoon and nighttime traffic as the main body, respectively \cite{Liangzhi}. Therefore a total of nine traffic components are required for a whole week. The abbreviations of these traffic components are listed in Table I.
\begin{table}
\begin{center}
\caption{The symbols corresponding to the 9 traffic components.}
\label{tab1}
\begin{tabular}{| c | c | c | c |}
\hline
 & Weekday & Saturday & Sunday\\
\hline
Morning & mw & msa & msu\\
\hline
Afternoon & aw & asa & asu\\
\hline
Evening & ew & esa & esu\\
\hline 
\end{tabular}
\end{center}
\end{table} 
Each traffic component is modeled as a Gaussian signal that can be expressed as

\begin{equation}
\setlength\abovedisplayskip{3pt}
\setlength\belowdisplayskip{3pt}
\label{deqn_ex2}
Y_{\rm c} = R_{\rm c}{\exp\left( {- \frac{\left( {t - t_{\rm c}} \right)^{\rm 2}}{{{\rm 2}\sigma}_{\rm c}^{\rm 2}}} \right)},
\end{equation}

\noindent where $ {t}_{\rm c} $ and $ {\sigma}_{\rm c}^{\rm 2} $ denote the mean value and variance of the Gaussian signal, respectively. $ R_{\rm c} $ is the peak value of the traffic component. Take Sunday as an example, the daily traffic can be represented as

\begin{equation}
\setlength\abovedisplayskip{3pt}
\setlength\belowdisplayskip{3pt}
\label{deqn_ex2}
Y_{\rm c_{1}}(t) ={\sum \limits_{\rm c_{1}} R_{\rm c_{1}}{\exp\left( {- \frac{\left( {t - t_{\rm c_{1}}} \right)^{\rm 2}}{{{\rm 2}\sigma}_{\rm c_{1}}^{\rm 2}}} \right)}},
\end{equation}

\noindent where $t$ is in a 24-hour format and ${\rm c_{1}} \in {\{ {\rm msu,asu,esu} \}}$. Therefore, the hourly traffic at time $t$, the $k$th day of a week, can be represented as

\begin{equation}
\setlength\abovedisplayskip{3pt}
\setlength\belowdisplayskip{3pt}
\label{deqn_ex6}
\begin{small}
\begin{matrix}
{{Y_{k}}(t) = {\sum\limits_{\rm c_{1}}^{}{{R_{\rm c_{1}}}{\sum\limits_{n_{\rm d} = 1}^{5}{\exp\left( {- \frac{\left( {t + 24\left( {n_{\rm d} - k} \right) - t_{\rm c_{1}} } \right)^{2}}{{2\sigma}_{\rm c_{1}}^{2}}} \right)}}}}} \\
{+ {\sum\limits_{\rm c_{2}}^{}{{R_{\rm c_{2}}}{\exp\left( {- \frac{\left( {t + 24\left( {6 - k} \right) - t_{\rm c_{2}} } \right)^{2}}{{2\sigma}_{\rm c_{2}}^{2}}} \right)}}}} \\
{+ {\sum\limits_{\rm c_{3}}^{}{{R_{\rm c_{3}}}{\exp\left( {- \frac{\left( {t + 24\left( {7 - k} \right) - t_{\rm c_{3}} } \right)^{2}}{{2\sigma}_{\rm c_{3}}^{2}}} \right).}}}} \\
\end{matrix}
\end{small}
\end{equation}

\noindent with the index of the day $k \in [1,2,3,4,5,6,7]$, the index of the weekday ${ n_{\rm d}}$, ${\rm c_{1}} \in {\{ {\rm mw,aw,ew} \}}$, ${\rm c_{2}} \in {\{{\rm msa,asa,esa}\}}$, and ${\rm c_{3}} \in {\{{\rm msu,asu,esu}\}}$.
The optimal parameter set of the daily traffic can be obtained by optimizing the following equation.

\begin{equation}
\setlength\abovedisplayskip{3pt}
\setlength\belowdisplayskip{3pt}
\label{deqn_ex7}
\underset{\begin{matrix}
{R_{\rm c},t_{\rm c},\sigma_{\rm c}^{2}} \\
{{\rm c} \in \left\{
{\begin{smallmatrix}
{\rm mw,aw,ew, }\\
{\rm msa,asa,esa, }\\
{\rm msu,asu,esu }\\
\end{smallmatrix}}
\right\}}
\end{matrix}}
{\rm minimize} \sum\limits_{k = 1}^{7} \sum\limits_{t = 1}^{24} \lVert {Y_{k}}(t) - Y_{\rm measure}(t) \rVert^{2},
\end{equation}

\noindent where $Y_{\rm measure}(t)$ refers to the traffic measurement at the moment $t$. This paper uses the gradient descent approach to tackle the minimization problem. 

\begin{figure*}[t]
\centering
\subfigure[Inter-Milan vs. Sampdoria kicks off at 16:00 December 1, 2013]{\begin{minipage}{3.4in}
    \includegraphics[width=1.1\textwidth]{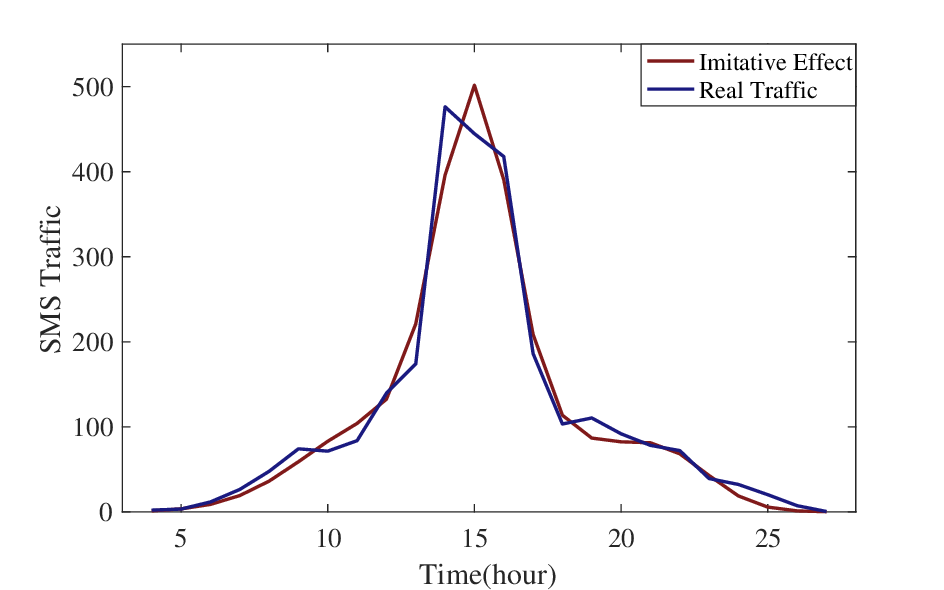}
\end{minipage}}
\subfigure[Inter-Milan vs. Trapani kicks off at 22:00 December 4, 2013]{\begin{minipage}{3.4in}
    \includegraphics[width=1.1\textwidth]{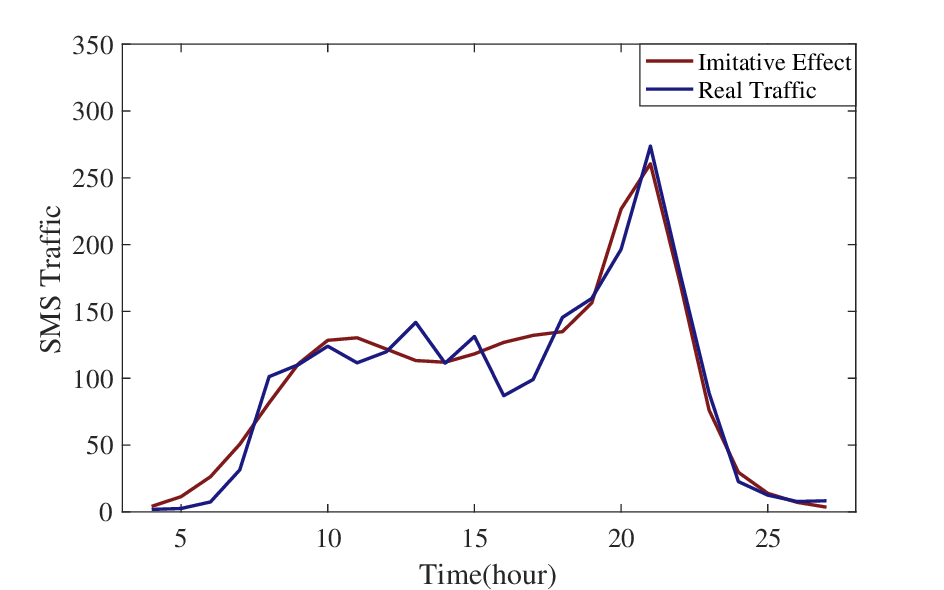}
\end{minipage}}
\subfigure[Inter-Milan vs. Parma kicks off at 21:45 December 8, 2013]{\begin{minipage}{3.4in}
    \includegraphics[width=1.1\textwidth]{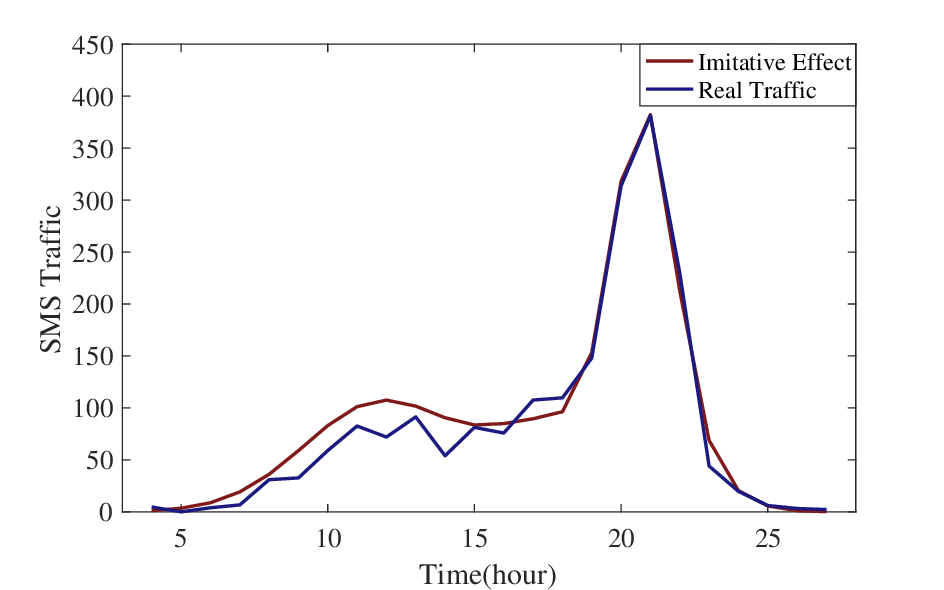}
\end{minipage}}
\subfigure[AC-Milan vs. Ajex kicks off at 21:45 December 11, 2013]{\begin{minipage}{3.4in}
    \includegraphics[width=1.1\textwidth]{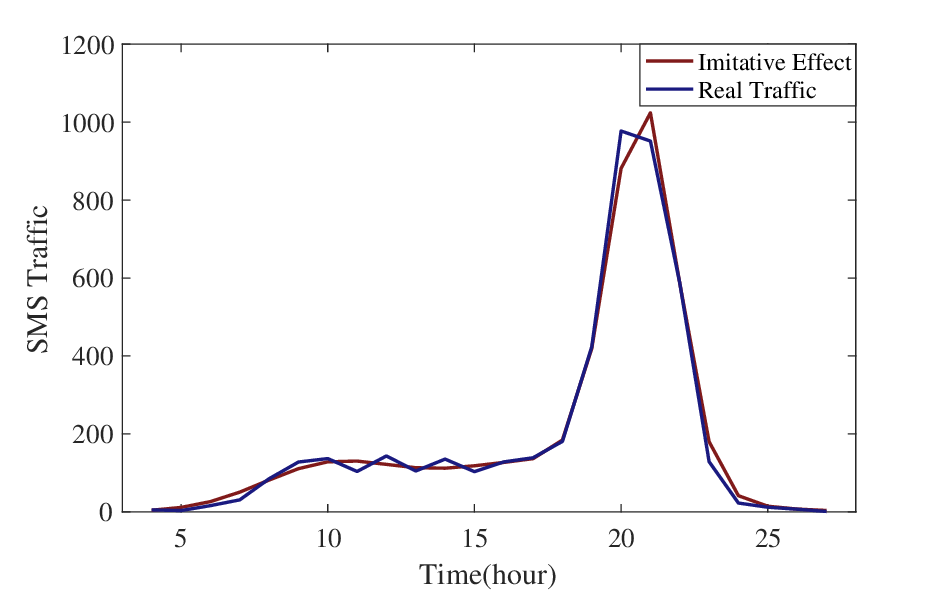}
\end{minipage}}
\label{fig5}
\caption{The imitative effect of SG-NNTP model for the practical traffic.}
\end{figure*}

According to the observation of Fig. \ref{SMS data for the soccer games}, it is obvious that the practical traffic data caused by the soccer games are similar to a bell-shaped curve. We assume that the behavior of attendances using cellular network services obeys an independent identically distribution, which is a logical general assumption. Then according to the central limit theorem, the prior distribution of the additive nonroutine traffic is naturally constructed as a Gaussian signal, which is consistent with the observation. Hence, the the additive nonroutine traffic can be represented as

\begin{equation}
\setlength\abovedisplayskip{3pt}
\setlength\belowdisplayskip{3pt}
\label{deqn_ex1}
Y_{\rm additive}(t) =P{\exp\left( {- \frac{\left( {t - t_{\rm sg}} \right)^{\rm 2}}{{{\rm 2}\sigma}_{\rm sg}^{\rm 2}}} \right)},
\end{equation}

\begin{equation}
\setlength\abovedisplayskip{3pt}
\setlength\belowdisplayskip{3pt}
\label{deqn_ex1}
P =\frac{R_{\rm sg}}{{\sigma_{\rm sg}\sqrt{2\pi}}},
\end{equation}

\noindent where the granularity of time $t$ is an hour. $ P $ is the peak value of the nonroutine traffic. $ R_{\rm sg} $ represent the amplitude parameter. ${t}_{\rm sg}$ and ${\sigma}_{\rm sg}^{\rm 2}$ are the mean and the variance of the Gaussian signal, respectively.
As shown in Fig. \ref{SMS data for the soccer games}, the horizontal coordinates of the peaks have a strong connection with the commencement time of the soccer games. 

\begin{table}
\begin{center}
\caption{The parameters corresponding to the first 4 soccer games in December.}
\label{tab1}
\begin{tabular}{| c | c | c | c | c |}
\hline
Date & Dec. 01 & Dec. 04 & Dec. 08 & Dec. 11\\
\hline
${t}_{\rm sg}$ & 15 & 21 & 20.75 & 20.75\\
\hline
${\sigma}_{\rm sg}$ & 1.263 & 1.176 & 1.011 & 1.155\\
\hline
$R_{\rm sg}$ & 829.8 & 349.8 & 769.7 & 2059.8\\
\hline
\end{tabular}
\end{center}
\end{table}

The traffic volume generated on the day when the soccer game occurs is equal to the daily component plus the additive nonroutine component, which is given by

\begin{equation}
\setlength\abovedisplayskip{4pt}
\setlength\belowdisplayskip{4pt}
\label{deqn_ex1}
{Y_{\rm total}(t)} = {Y_{\rm daily}(t)} + {Y_{\rm object}(t)},
\end{equation}

\noindent where ${Y_{\rm object}(t)}$ is ideal for the additive nonroutine component at the moment $t$.
Since the daily traffic component ${Y_{\rm daily}(t)}$ can obtained by UBBNTP method as mentioned above, ${Y_{\rm object}(t)}$ can be obtained according to formula (7). The least squares method is then used to minimize the difference between ${Y_{\rm object}}$ and ${Y_{\rm additive}}$, which is formulated as

\begin{equation}
\setlength\abovedisplayskip{4pt}
\setlength\belowdisplayskip{4pt}
\label{deqn_ex7}
\underset{\begin{matrix}
{R_{\rm sg}, {t}_{\rm sg}, \sigma_{\rm sg}^{2}}
\end{matrix}}
{\rm minimize} \sum \lVert {Y_{\rm additive}(t)} - Y_{\rm object}(t) \rVert^{2},
\end{equation}

\noindent and thus, the parameters, i.e. ${t}_{\rm sg}$, $R_{\rm sg}$, and ${\sigma}_{\rm sg}$ corresponding to each soccer game, are obtained.

Fig. 5 illustrates the fitting performance of the SG-NNTP model for the first 4 soccer games in December 2013 at San Siro stadium, Milan. 
Approximating $ {t}_{\rm sg} $ to the commencement time of the soccer games simplified the problem and led to successful outcomes. The parameters of the simulation are shown in Table II.

\subsection{Multistep Prediction of SG-NNTP}
One of the major advantages of the analytical model is that it can efficiently and accurately perform multi-step predictions. It requires the ability to estimate the initial parameters of the SG-NNTP model in advance based on the advanced information of the event. The first four soccer games hosted by the San Siro Stadium in December 2013 are then used as a training set in this paper to derive the relationships between attendance and the initial parameters.

\begin{figure}[!t]
\centering 
\includegraphics[width=3.4in]{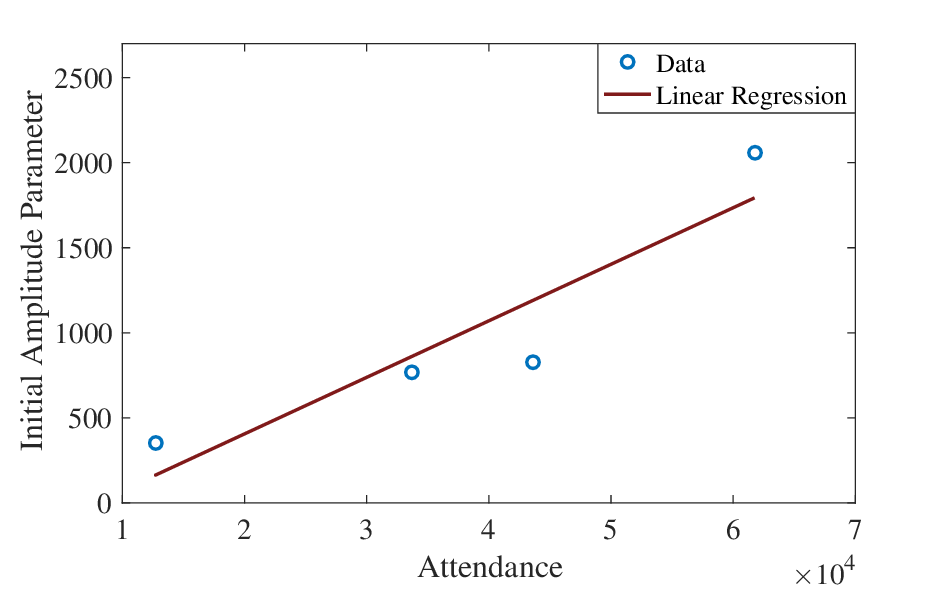}
\caption{Linear regression between $ R_{\rm sg} $ and attendance.}
\label{attendance and the initial parameters}
\end{figure}

The data volume of additive nonroutine traffic can be expressed as the integral of $ Y_{\rm additive} $ over $ (- \infty, + \infty) $, exactly as $ R_{\rm sg} $, which is given by

\begin{equation}
\setlength\abovedisplayskip{3pt}
\setlength\belowdisplayskip{3pt}
\label{deqn_ex1}
\int_{- \infty}^{+ \infty} \frac{R_{\rm sg}}{{\sigma_{\rm sg}\sqrt{2\pi}}}{\exp\left( {- \frac{\left( {t - t_{\rm sg}} \right)^{\rm 2}}{{{\rm 2}\sigma}_{\rm sg}^{\rm 2}}} \right)} = R_{\rm sg}.
\end{equation}

The Pearson correlation coefficient $r$ between initial parameter $R_{{\rm sg}_j}$ and attendance $ A_{j} $ is calculated as

\begin{equation}
\setlength\abovedisplayskip{3pt}
\setlength\belowdisplayskip{5pt}
\label{deqn_ex8}
r = \frac{\sum_{j=1}^{n}({R_{{\rm sg}_j}} - \bar{R})(A_{j} - \bar{A})}{\sqrt{\sum_{j=1}^{n}({R_{{\rm sg}_j}} - \bar{R})^2} \cdot \sqrt{\sum_{j=1}^{n}(A_{j} - \bar{A})^2}},
\end{equation}

\begin{figure}[t]
\centering
\subfigure[the \textbf{proposed} SG-NNTP model]{\begin{minipage}{3.4in}
    \includegraphics[width=1\textwidth]{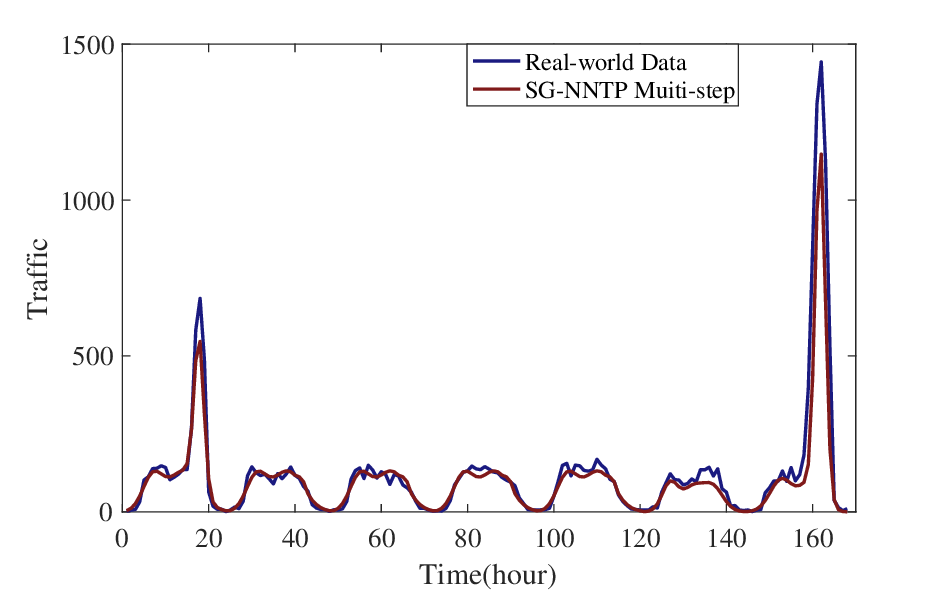}
\end{minipage}}
\subfigure[the 5-layer MLP model]{\begin{minipage}{3.4in}
    \includegraphics[width=1\textwidth]{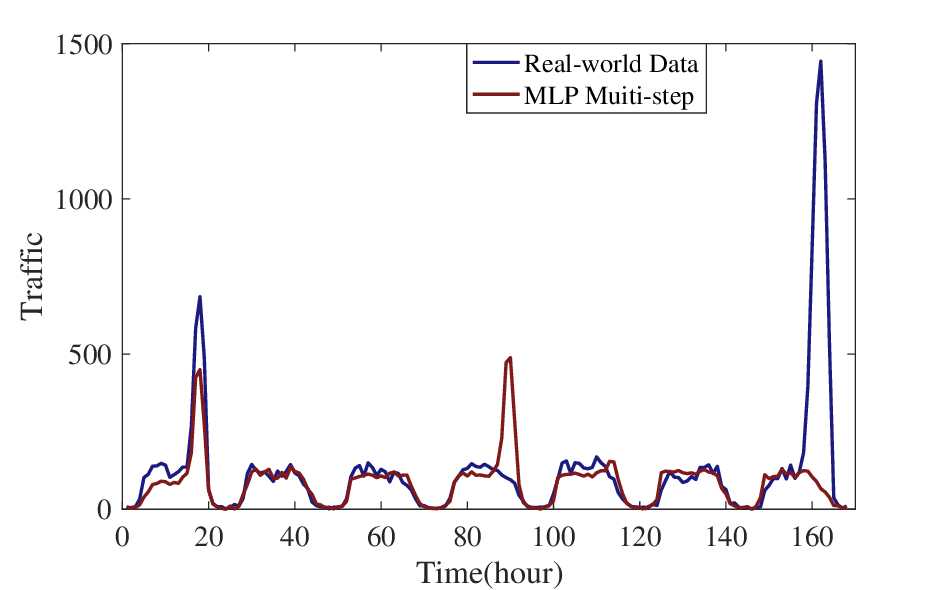}
\end{minipage}}
\subfigure[the 5-layer LSTM model]{\begin{minipage}{3.4in}
    \includegraphics[width=1\textwidth]{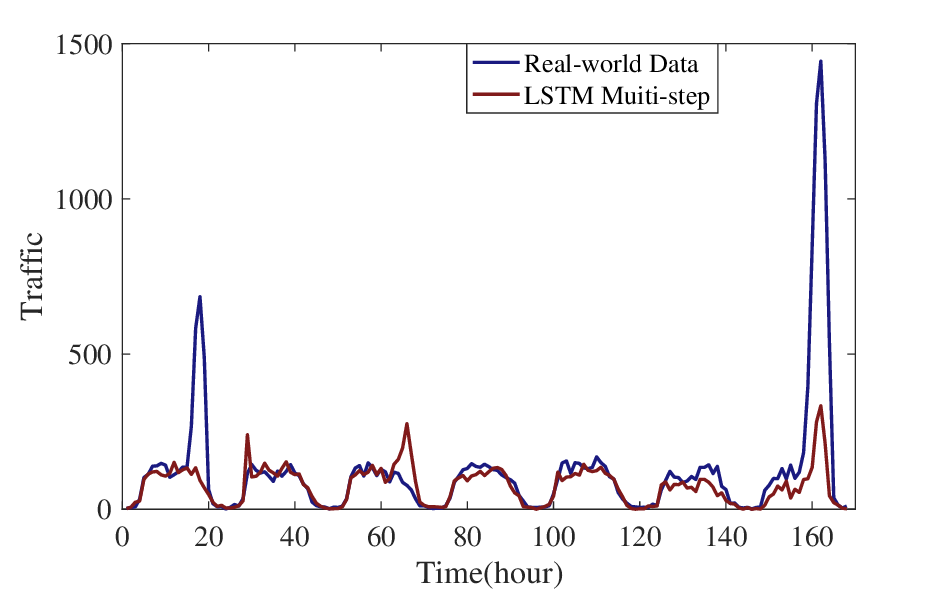}
\end{minipage}}
\label{The imitative effect of SG-NNTP model}
\caption{Prediction results for SG-NNTP model and the benchmark models in multi-step prediction mode.}
\end{figure}

\begin{figure}[!t]
\centering 
\includegraphics[width=3.4in]{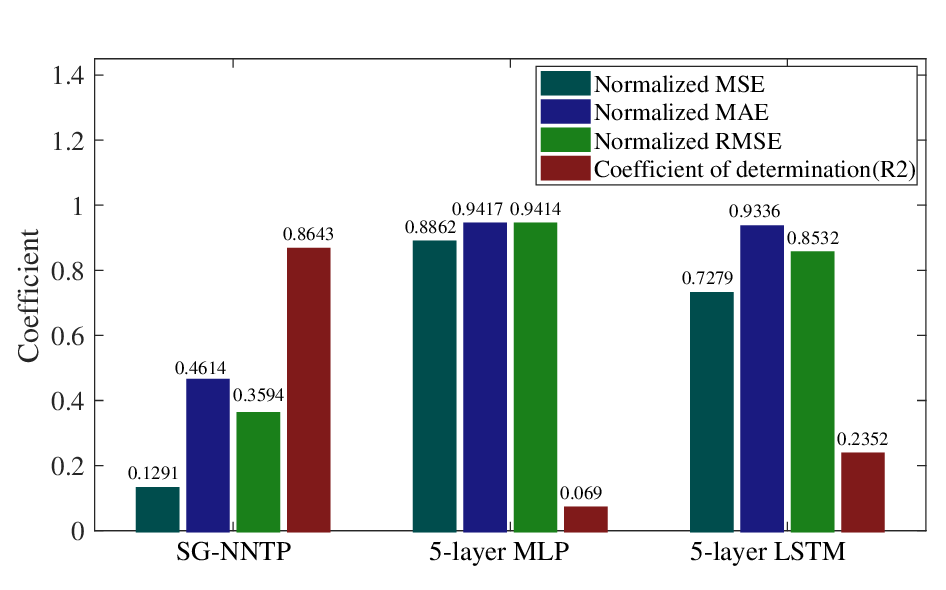}
\caption{The accuracy of the SG-NNTP model and the benchmark models in multi-step prediction mode.}
\label{fig8}
\end{figure}

\noindent where $n = 4$ denotes the four soccer games, ${R_{{\rm sg}_j}}$ and $A_{j}$ denote the attendance and initial amplitude parameters corresponding to the $j$th soccer game, respectively. After calculation, the value of $r$ is 92.2\%, which means there is a strong positive correlation between $ R_{\rm sg} $ and attendance. With a sample size of only 4, there is little benefit in constructing complicated correspondence between $ R_{\rm sg} $ and attendance. Hence, this paper applies the simplest linear regression approach to determine the correspondence as shown in Fig. \ref{attendance and the initial parameters}.
The correspondence is expressed as

\begin{equation}
\setlength\abovedisplayskip{3pt}
\setlength\belowdisplayskip{3pt}
\label{deqn_ex8}
{R_{{\rm sg}_j}} = A_{j} \times 0.03323 - 258.85.
\end{equation}

The parameter ${\sigma}_{\rm sg}$ characterizes the habits of all attendance in using cellular network traffic during the soccer game period. 
${\sigma}_{\rm sg}$ has a complex relationship with external information such as attendance, type of game, the popularity, the intensity, etc.
The implicit relationship between ${\sigma}_{\rm sg}$ and external information cannot be fully corroborated due to the small amount of samples. Therefore, this paper adopts the average value to represent ${\sigma}_{\rm sg}$  of SG-NNTP model. With the ability to estimate the initial parameters, consequently our model is able to perform multi-step prediction.

\begin{figure}[!t]
\centering 
\includegraphics[width=3.4in]{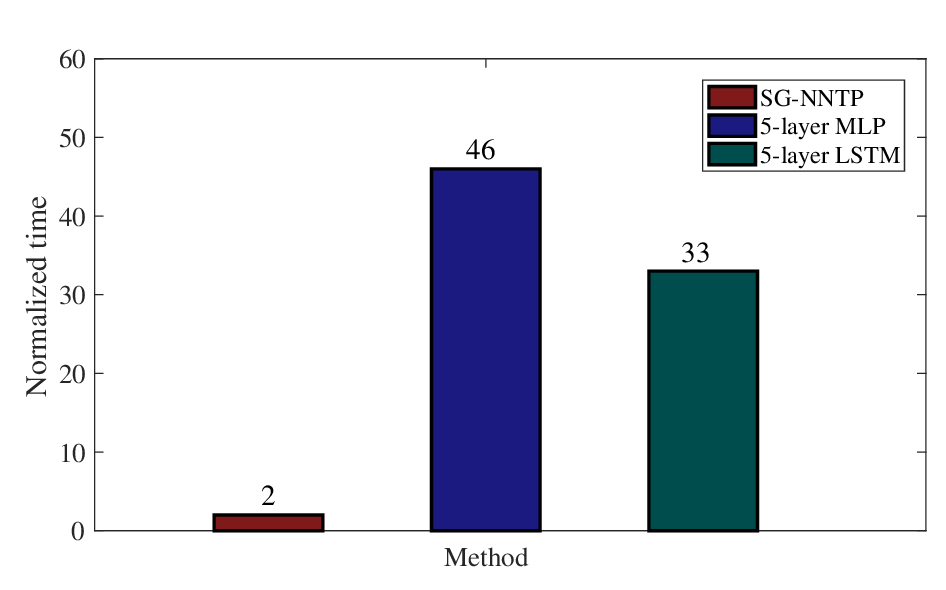}
\caption{The elapsed time of the SG-NNTP model and the benchmark models in multi-step prediction mode.}
\label{fig9}
\end{figure}

\subsection{Single-step Prediction of SG-NNTP}
The single-step mode predicts the forthcoming traffic value by utilizing the actual traffic data at the present moment.
Compared to multi-step prediction that focuses on the overall trend, single-step prediction is only concerned with the traffic value at the next moment. With the support of real time data entered at each time step, single-step prediction produces more accurate results. Therefore, both ML-based and statistics-based NTP model in the mainstream adopt single-step prediction.

The proposed SG-NNTP model also possesses the capability to execute single-step prediction, thereby achieving outstanding performance.
The parameters of the model are constantly updated based on real-time practical input data. We adopt the least squares method to minimize the following equation to achieve the update of the parameters, which is formulated as

\begin{equation}
\setlength\abovedisplayskip{3pt}
\setlength\belowdisplayskip{3pt}
\label{deqn_ex7}
\underset{\begin{matrix}
{R_{{\rm sg}_{i}},\sigma_{{\rm sg}_i}^{2}}
\end{matrix}}
{\rm minimize} \sum\limits_{t_1}^{t_i} \lVert {Y_{additive}(t_i)} - {Y_{\rm object}(t_i)} \rVert^{2}.
\end{equation}

\noindent Then $R_{{\rm sg}_{i}}$ and $\sigma_{{\rm sg}_i}$ are used to predict the traffic at the moment $t_{i+1}$, where the predicted values at the initial moment are obtained from the initial parameters.

In addition, to reduce the impact of initial parameters on the prediction performance, a least squares optimization method with multiple initial values has been adopted. In each step of the update, multiple sets of initial parameters are selected for optimization and the smallest MSE counterpart is used for predicting the next moment traffic.

\section{Evaluation with Real-world Traffic Data}
This section primarily evaluates the prediction accuracy and computational efficiency of the proposed SG-NNTP model in comparison to benchmark models.
The evaluation indexes of prediction accuracy are Mean Square Error (MSE), Root MSE (RMSE), Mean Absolute Error (MAE), and R-squared coefficient (R2), which are defined as

\begin{equation}
\label{deqn_ex8}
{{\rm MSE}} = \frac{1}{n}\sum_{i=1}^n(y_i - \hat{y}_i)^2,
\end{equation}

\begin{equation}
\label{deqn_ex9}
{{\rm RMSE}} = \sqrt{{\rm MSE}},
\end{equation}

\begin{equation}
\label{deqn_ex10}
{{\rm MAE}} = \frac{1}{n}\sum_{i=1}^n\left|y_i - \hat{y}_i\right|,
\end{equation}

\begin{equation}
\label{deqn_ex11}
{{\rm R2}} = 1 - \frac{\sum_{i=1}^n(y_i - \hat{y}_i)^2}{\sum_{i=1}^n(y_i - \bar{y})^2},
\end{equation}

\noindent where $n$ represents the number of predicted samples, $y_i$ represents the actual value, $\hat{y}_i$ represents the predicted value, and $\bar{y}$ denotes the mean value of $y_i$.
Computational efficiency is represented by the elapsed time including training and prediction time. 

\begin{figure*}[t]
\centering
\subfigure[the 5-layer MLP model]{\begin{minipage}{3.4in}
    \includegraphics[width=1\textwidth]{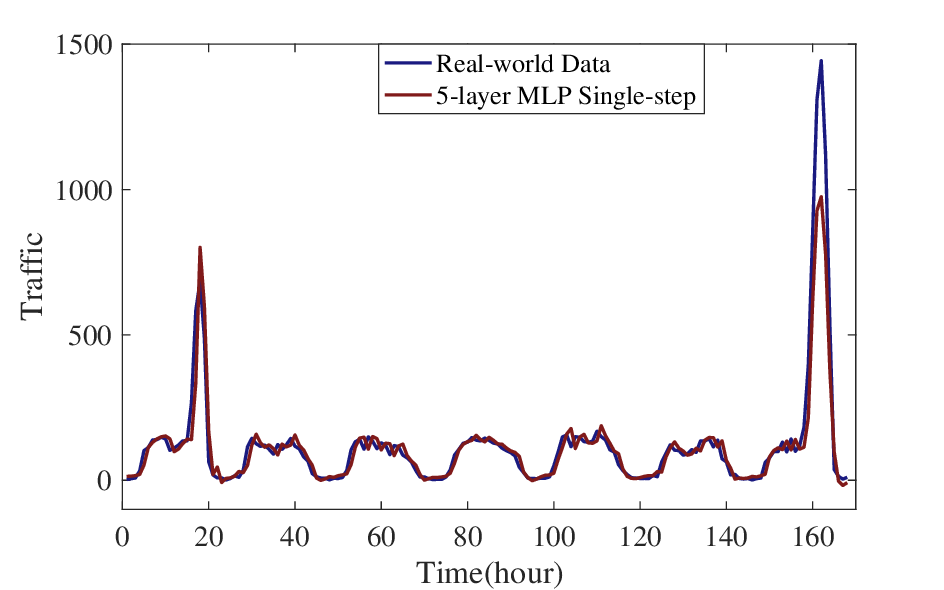}
\end{minipage}}
\subfigure[the 3-layer LSTM model]{\begin{minipage}{3.4in}
    \includegraphics[width=1\textwidth]{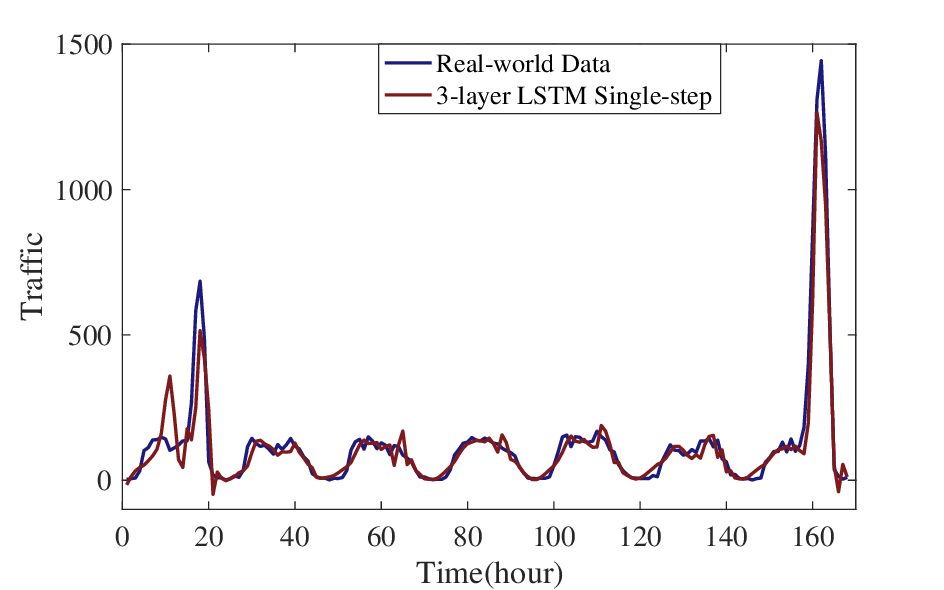}
\end{minipage}}
\subfigure[the ARMA(1,2) model]{\begin{minipage}{3.4in}
    \includegraphics[width=1\textwidth]{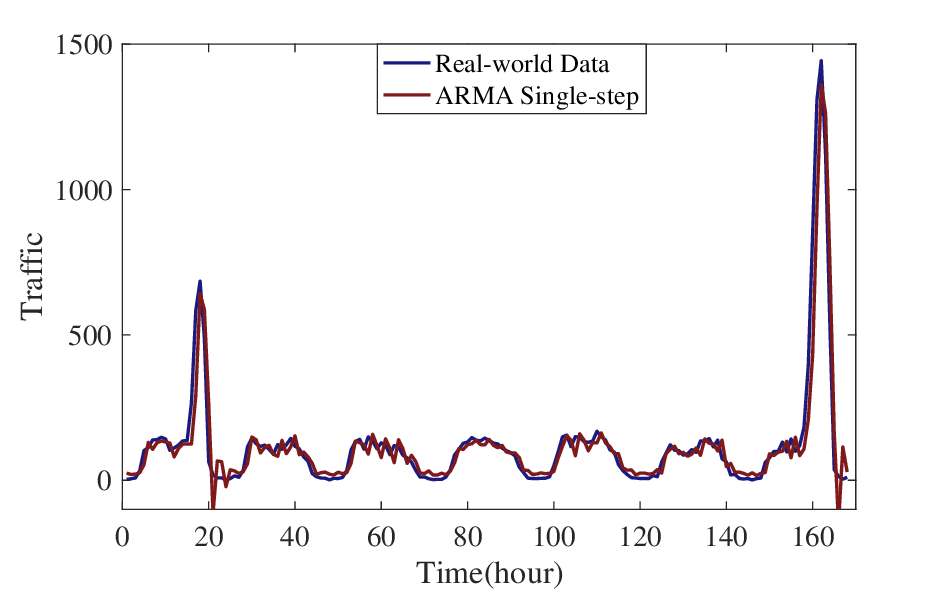}
\end{minipage}}
\subfigure[the ARIMA(3,1,1) model]{\begin{minipage}{3.4in}
    \includegraphics[width=1\textwidth]{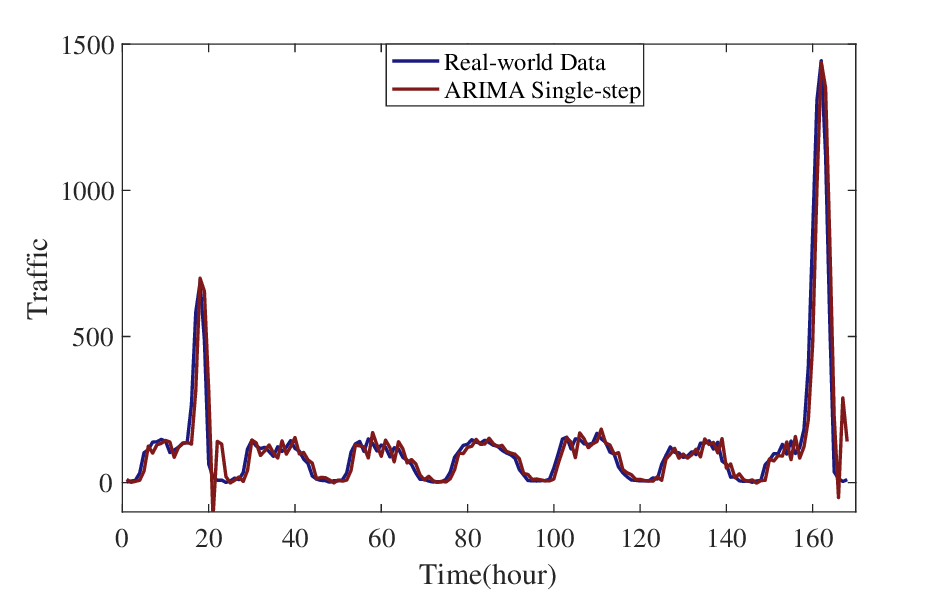}
\end{minipage}}
\caption{Prediction results for the benchmark models in single-step prediction mode.}
\end{figure*}

\begin{figure}[!t]
\centering 
\includegraphics[width=3.4in]{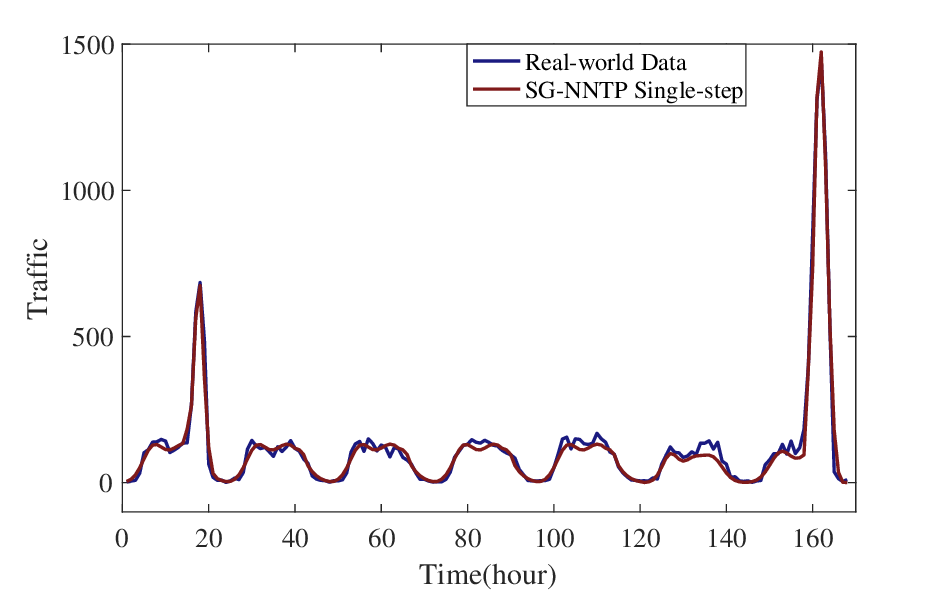}
\caption{Prediction results for the proposed SG-NNTP model.}
\label{fig11}
\end{figure}

Deep Multi-Layer Perceptron (MLP) network and classic recurrent neural networks (i.e. LSTM network), are chosen to construct the benchmark models which represent the ML-based NTP model. Meanwhile, the ARMA and ARIMA models are also used to represent the classic statistics-based NTP model. According to Bayesian Information Criterion (BIC), the parameters of ARMA and ARIMA models are determined to be ARMA (1,2) and ARIMA (3,1,1). Traffic data from December 1st to 15th is used as the training set for all models, and the data from December 16th to 22th is used as the test set. As shown in Fig. \ref{soccer games information}, the training set contains four pieces of additive nonroutine traffic corresponding to four soccer games and the test set contains two.

\subsection{Performance Comparison Between the SG-NNTP Model and the Benchmark Models in Multi-step Prediction Mode}
The capacity to predict many time steps reflects NTP models' understanding of overall trends of cellular network traffic. 
The longer the processing step, the higher the level of difficulty, and the higher the likelihood that prediction accuracy will be compromised.
In this paper, the time step is set to one hour. When the prediction step size is set to 168 time steps, i.e., directly predicting the traffic for a whole week, the proposed SG-NNTP model exhibits excellent performance in terms of prediction accuracy with the R2 coefficient of 86.4\%, as shown in Fig. 7 (a). As benchmark models, this paper selects 5-layer MLP network and 5-layer LSTM network to represent ML-based NTP model, and the ARMA and ARIMA models to represent statistics-based models. Experimental results demonstrate that these benchmark models can not effectively extract the traffic pattern when the prediction step size is too long. 
Therefore, we lower the level of difficulty by shortening the prediction step size of the benchmark models to 24 time steps. At this point, the statistics-based ARMA and ARIMA models still cannot work well. 
While ML-based models are superior to statistics-based models, it is significantly inferior to the SG-NNTP model.  
The prediction results of the MLP and LSTM networks are shown in Fig. 7 (b) and (c), respectively.

As can be seen from Fig. 7, in the presence of the nonroutine event, the benchmark models can not effectively extract the traffic pattern and understand the overall trend of network traffic from the historical data. In the case of reduced prediction difficulty of benchmark models, i.e., shorter prediction step size, the benchmark models are still far inferior to the proposed SG-NNTP model, both in terms of prediction accuracy and computational efficiency. As shown in Fig. \ref{fig8}, the SG-NNTP model achieves the highest R2 which is about 3.7 to 12.5 times higher than the R2 of the benchmark models. Meanwhile, the proposed SG-NNTP model decreases the MSE, MAE, and RMAE coefficients by 82.3\%, 50.6\%, and 57.9\%, respectively, in comparison to the LSTM models which is the best performing of the benchmark models. From the perspective of computational efficiency, as shown in Fig. \ref{fig9}, our model is 16.5 times more efficient than the 5-layer LSTM network and 23 times more efficient than the 5-layer MLP network.

\begin{figure}[!t]
\centering 
\includegraphics[width=3.4in]{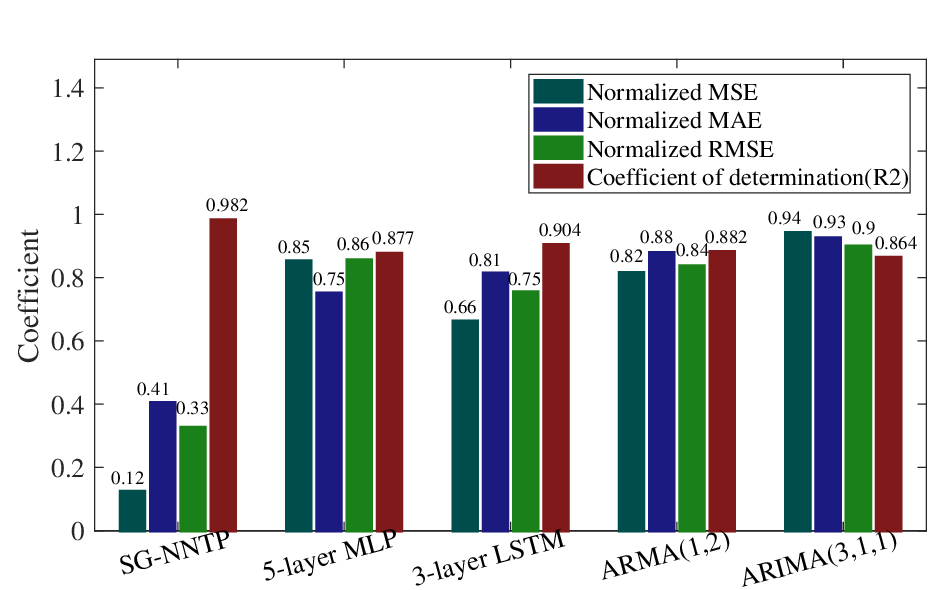}
\caption{The accuracy of the SG-NNTP model and the benchmark models in single-step prediction mode.}
\label{fig12}
\end{figure}

\subsection{Performance of the SG-NNTP Model and Benchmark Models in Single-step Prediction Mode}
The state-of-the-art regarding NTP mainly focuses on single-step prediction mode. This mode reflects NTP model's ability to capture localized characteristics of the traffic data. NTP models which adopt single-step prediction mode tend to have higher accuracy, due to the support of real-time traffic data at each time step. As shown in Figs. 10 and \ref{fig11}, both the SG-NNTP model and the benchmark models show an intuitive improvement in terms of prediction accuracy when adopting the single-step prediction mode.

\begin{figure}
\centering 
\includegraphics[width=3.4in]{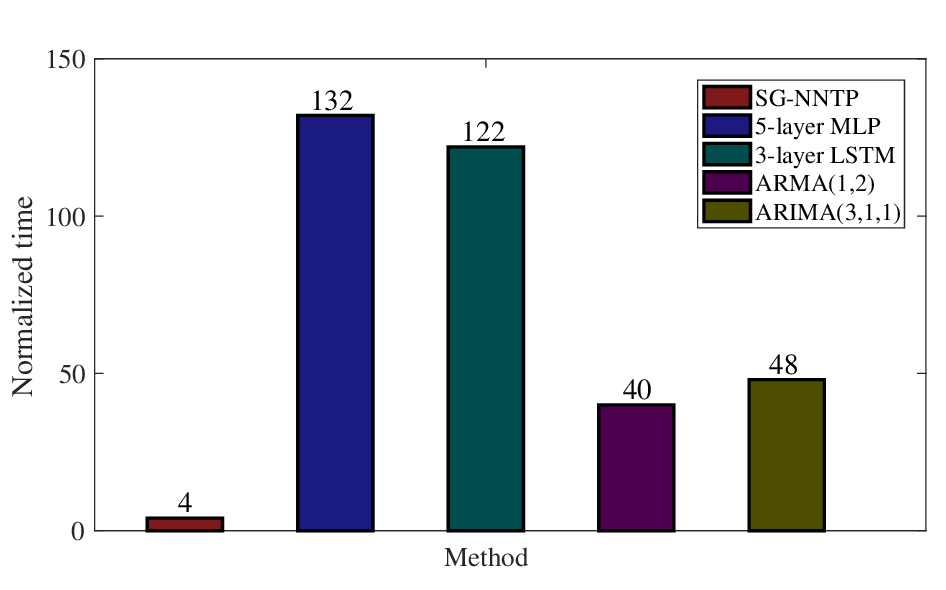}
\caption{The elapsed time of the SG-NNTP model and the benchmark models in single-step prediction mode.}
\label{fig13}
\end{figure}

Figs. 10 and \ref{fig11} represent the prediction results of the benchmark models and the proposed SG-NNTP model. 
In comparison with the benchmark models shown in Fig. 10, it is obviously that the proposed SG-NNTP model achieves prominent advantage in prediction accuracy as shown in Fig. \ref{fig11}.
More intuitively, Fig. \ref{fig12} quantifies their performance with the evaluation indexes of prediction accuracy. 
Among the benchmark models, the LSTM network achieves the best performances in terms of MSE, RMSE, and R2 coefficient, while the MLP network possesses the minimum MAE. 
However, there is still a considerable gap between the benchmark models and the SG-NNTP model. 
As shown in Fig. \ref{fig12}, the SG-NNTP model achieves the R2 coefficient as high as 98.2\%, while decreasing the the MSE and RMSE coefficients by 81.8\% and 56\%, respectively, in comparison to the LSTM network. Furthermore, the SG-NNTP model reduces the MAE coefficient by 45.3\% compared with the MLP network. 
Fig. \ref{fig13} demonstrates the performances of all models in terms of elapsed time. Our model also has the highest computational efficiency which is about 33 times that of the MLP network and about 31 times that of the LSTM network as shown in Fig. \ref{fig13}.

\subsection{Analysis and Discussion}
Due to the diversity of event types, occurrence times, durations, and user behaviors during events, NTP models built entirely on historical traffic data suffer from low prediction accuracy, computational efficiency, and interpretability. In contrast, the prediction accuracy and computational efficiency of the proposed SG-NNTP model outperform those of the benchmark models, both in single-step prediction mode and multi-step prediction mode. Meanwhile, the SG-NNTP model is an analytical NTP model constructed according to the novel NNTP method based on the analysis of user behavior. Therefore, this model possesses highly interpretability. As with all analytical models, the model parameters are concise, and practically meaningful, which can be efficiently and accurately applied to similar events. Mutually, excellent prediction performance and generalization abilities also indirectly proves that the novel NNTP method is reasonable and consistent with network traffic pattern.

More importantly, the NNTP method pioneers the correspondence between infrequent events and specific NTP model, provides an approach to analysis and process the network traffic caused by nonroutine events, and is an inspiration for subsequent research on nonroutine traffic. In addition, analyzing the causes of formation and trends of nonroutine traffic can be very helpful for accurate and efficient network resource allocation.

What is more, the proposed NNTP method is essentially an efficient synthesis of multivariate data, which makes a recommendation for data collection and storage in the context of cellular network traffic. Specifically, additional information, such as regional events, number of users, etc, could be collected and stored with traffic data, which will be very beneficial for future research of nonroutine traffic. Similar events can potentially be further subdivided based on more detailed event information. In this way, more accurate NTP models with more reliable implicit relationships can be discovered. In fact, it is found that there exists some implicit relation between the parameter $\sigma_{sg}$ and the number of attendances for all soccer games belonging to the Italian Serie A type. This relation is really helpful to initial parameter estimation and enable the model to achieve more accurate multi-step prediction results. However, due to the limitation of sample size, the implicit relationship has not been fully validated and therefore was not applied in current SG-NNTP model.

\section{Conclusion}
This paper raised the problem about nonroutine traffic, and pioneered a novel NNTP method to analyze and process nonroutine traffic. Subsequently, this paper constructed the SG-NNTP model for the additive nonroutine traffic caused by soccer games as a case study to validate the performance of the NNTP method. Experimental results show that the NNTP method outperforms the benchmark models in prediction accuracy, both in single-step and multi-step prediction mode. Also, the computational efficiency is greatly improved. In addition, the model constructed by the NNTP method is an analytical NTP model based on user behaviour analysis with outstanding interpretability.

\end{document}